\documentstyle[12pt]{article}
\begin{document}
\def\om{\omega}
\def\omt{\tilde{\omega}}
\def\ti{\tilde}
\def\o{\Omega}
\def\t{T^*M}
\def\vt{\tilde{v}}
\def\ot{\tilde{\Omega}}
\def\otwo{\omt \wedge \om}
\def\owot{\om \wedge \omt}
\def\w{\wedge}
\def\mt{\tilde{M}}
\def\T{{\rm Tr}}
\def\om{\omega}
\def\omt{\tilde{\omega}}
\def\ss{\subset}
\def\bc{{\bf C}}

\def\om{\omega}
\def\omt{\tilde{\omega}}
\def\ti{\tilde}
\def\o{\Omega}
\def\t{T^*M}
\def\vt{\tilde{v}}
\def\ot{\tilde{\Omega}}
\def\otwo{\omt \wedge \om}
\def\owot{\om \wedge \omt}
\def\w{\wedge}
\def\mt{\tilde{M}}

\def\om{\omega}
\def\omt{\tilde{\omega}}
\def\ss{\subset}
\def\tpm{T_{P} ^* M}
\def\al{\alpha}
\def\alt{\tilde{\alpha}}
\def\la{\langle}
\def\ra{\rangle}
\def\inop{{\int}^{P}_{P_{0}}{\om}}
\def\th{\theta}
\def\lm{\lambda}
\def\tht{\tilde{\theta}}
\def\inox{{\int}^{X}{\om}}
\def\inotx{{\int}^{X}{\omt}}
\def\st{\tilde{S}}
\def\ls{\lambda_{\sigma}}
\def\p{{\bf{p}}}
\def\pb{{\p}_{b}(t,u)}
\def\pbm{{\p}_{b}}
\def\d{\partial}
\def\d+{\partial_+}
\def\d-{\partial_-}
\def\pat{\partial_{\tau}}
\def\pas{\partial_{\sigma}}
\def\dpm{\partial_{\pm}}
\def\l2{\Lambda^2}
\def\be{\begin{equation}}
\def\ee{\end{equation}}
\def\bea{\begin{eqnarray}}
\def\eea{\end{eqnarray}}
\def\ej{{\bf E}}
\def\ed{{\bf E}^\perp}
\def\si{\sigma}
\def\cg{{\cal G}}
\def\cgt{\ti{\cal G}}
\def\cd{{\cal D}}
\def\r{{\cal R}}
\def\ce{{\cal E}}
\def\cep{\ce^{\perp}}
\def\cf{{\cal F}}
\def\cfp{\cf^{\perp}}
\def\bz{\bar{z}}
\def\e{\varepsilon}
\def\b{\beta}
\begin{titlepage}
\begin{flushright}
{}~
CERN-TH/96-254\\
hep-th/9609112
\end{flushright}

\vspace{1cm}
\begin{center}
{\Large \bf Open strings and $D$-branes in WZNW models}\\
[50pt]{\small
{\bf C. Klim\v{c}\'{\i}k}\footnote{Address after 1 October  1996: IHES, 
91440 Bures-sur-Yvette, France}\\
Theory Division, CERN,\\ 
CH-1211 Geneva 23, Switzerland \\[3pt] and \\[3pt] 
{\bf P. \v Severa }\\
Department of Theoretical Physics, Charles University, \\
V Hole\v sovi\v ck\'ach 2, CZ-18000 Praha,
Czech Republic\\[30pt] }

\begin{abstract}

An abundance of the Poisson-Lie symmetries of the  WZNW models
is uncovered. They give rise, via the Poisson-Lie $T$-duality, to a
rich structure of the dual pairs of $D$-branes configurations  in
group manifolds. The $D$-branes are characterized by their shapes and
certain two-forms living on them. The WZNW path integral for 
the interacting $D$-branes diagrams  is unambiguously defined if the two-form 
on the $D$-brane and 
the WZNW three-form on the group form an integer-valued 
 cocycle in the relative singular 
cohomology of the group manifold with respect to its $D$-brane submanifold.
An example of the $SU(N)$ WZNW model is studied in some detail.
\end{abstract}
\end{center}
\vskip 1.5cm
\noindent CERN-TH/96-254\\
September 1996

\end{titlepage}

 The Poisson--Lie (PL) $T$-duality \cite{KS2} is a generalization of the 
traditional non-Abelian $T$-duality \cite{OQ}--\cite{GR} and it  proved
to enjoy 
\cite{KS2},
\cite{T}--\cite{KS6}, at least at the classical level,  all of the
structural features of the traditional Abelian $T$-duality 
\cite{SS} and \cite{Busch}.  In particular, our so far last  paper on the
subject \cite{KS6}
has settled (at the classical level)  the  remaining big issue
of the PL generalization: the momentum-winding exchange.

It is now of an obvious interest to promote the PL $T$-duality to the
quantum world. Strictly speaking, a consistent quantum picture
does not necessarilly imply that mutually  dual quantum models have to
be conformally invariant. However,  we do wish to have {\it conformal} 
examples in order to apply the PL $T$-duality in string theory. In this paper
we shall show that such conformal examples of PL dualizable $\sigma$-models
are the standard WZNW models and we shall give the detailed classical
account of the PL $T$-duality for them. The
treatment of the first quantized strings 
 we postpone to a forth-coming publication,
where an  emergence of a proliferation of quantum group structures seems
unavoidable.

In what follows, we shall demonstrate that a PL dualizable
$\sigma$-model satisfying only a certain mild algebraic condition 
is necessarily a WZNW model. This means  that the WZNW models are 
not only `some' conformal examples of the 
 dualizable models but, in a sence, they are very characteristic
for the structure of the PL $T$-duality.  Moreover, for various Drinfeld 
doubles
underlying the structure of PL $T$-duality one recovers the same WZNW
model! 
  Hence, there are  many (in fact infinitely many) Poisson-Lie
symmetries in WZNW models. 

 It turns out that the dual to the WZNW model  is again the 
same WZNW model. This should not be interpreted as a drawback. After all, 
what really matters is the fact that this (self)-duality induces 
a  non-trivial  non-local map on the phase space of the model
which, in particular, reshuffles zero modes of the string, much in the
same way as in the Abelian $T$-duality. The fundamental groups 
of the compact non-Abelian groups\footnote{We wish to consider
the compact groups for  the string theory compactifications.} are rather
small therefore the momentum-winding exchange for  closed strings 
 may be rather modest (cf. \cite{KS6}). On the other hand,
the duality transformation of the zero modes  of  open strings gives
the rich and spectacular structure in the dual: the celebrated $D$-branes
\cite{Pol}.

We have devoted one paper in our series to the PL $T$-duality between
open strings and $D$-branes \cite{KS4}. It describes the geometries of 
the $D$-branes
for arbitrary perfect\footnote{Every  element of the `perfect' 
 Drinfeld double 
can be uniquely written as the product of two elements of the two
 groups forming the double. Since the semi-Abelian Drinfeld doubles,
that correspond to the traditional non-Abelian duality, are perfect
we did give the complete picture of the $T$-duality between open strings
and $D$-branes in this traditional case. Later 
also works 
by two different groups \cite{FKS,BL} appeared,  dealing with open strings in
traditional non-Abelian duality.}
Drinfeld double in terms of the symplectic leaves of the associated
Poisson homogeneous spaces \cite{KS4}.
In this contribution, we have to describe the open strings -  $D$-branes
duality also for non-perfect doubles in order to account for the WZNW
models. We shall again obtain a rich geometry of the $D$-branes
dictated by a simple structure on the double. We wish to stress at this
point that the PL $D$-branes are very different from the standard 
Abelian $D$-branes. In the latter case the $D$-branes are just
points in the direction of the space-time coordinates with respect to which
one performs the Abelian duality and they become  extended objects
only in the direction of the extra (spectator or  Buscher) duality intact
coordinates. In the PL case, however, the $D$-branes are not points
even without extending the space-time by the spectator coordinates!
They may posses quite a complicated geometry, as we shall see later on.

Our presentation will contain also the case of  open strings
in WZNW models. This is generally not a well defined system because it requires
a choice of the two-form potential of the WZNW three-form. This potential
is not only ambiguous but it must be also singular because the 
WZNW three-form is a nontrivial element of the third de Rham cohomology  
on the (compact) group manifold and, as such, it does not admit a globally 
defined
potential. It may therefore seem that there is a lot of arbitrariness
in defining  open strings in WZNW models. One has to choose the singular
points of the two-form potential and  the potential itself with the
condition that the dynamics will disallow the end-points of strings to hit the 
 singularity\footnote{The bulk of the string feels only the exterior
derivative of the two-form potential which is nothing but the perfectly regular
WZNW three-form.}. 

Some CFT results have been already obtained for open strings in $SU(2)$ WZNW 
model
in \cite{Sag}, however, we did not find a discuusion of the subtle issue of 
the meaning of the WZNW term for open
strings.
 In our case the arbitrariness in defining the WZNW model
for open strings is completely fixed by the 
requirement of the PL symmetry.
 By picking up  one of the Drinfeld doubles corresponding
to a given WZNW model and by fixing one half-dimensional 
isotropic subalgebra in the algebra
of the double, we   fix uniquely the singular two-form potential of the
WZNW three-form  and ensure that the end-points of strings do not hit the 
singular points
on the target. Moreover, there exists a dual $D$-brane configuration 
and its geometry is again given in terms of the simple data on the double.

In the first section of this paper,  we  provide a topological discussion of 
conditions
 when the WZNW path integral is well
defined for interacting string diagrams corresponding
to a given  $D$-brane configuration. Then we give
the description of the  Poisson-Lie  symmetries occuring in the  
WZNW models in terms of the  underlying Drinfeld doubles. In the third section
we  describe the $D$-brane configurations for a particular PL symmetry
also for the non-perfect doubles and describe the classical phase space of the
system.
We also formulate an easy {\it non}-cohomological criterion
when  the underlying data on the double give  a well defined
WZNW path integral for the interacting $D$-branes diagrams.
In the fourth  section we  provide  examples of the general construction:
PL symmetries and the $D$-branes in the $SU(N)$ WZNW models.

\section{$D$-branes and the WZNW path integral.}

The standard WZNW action on a group manifold $R$ 
 reads
\be S(r)\equiv {1\over 4\pi}\int d\xi^+ d\xi^-\la \partial_+ r~r^{-1}, 
\partial_- r~r^{-1}\ra +{1\over 24\pi} 
\int d^{-1}\la dr~r^{-1},[dr~r^{-1},
dr~r^{-1}]\ra .\ee
Here $\xi^{\pm}$ are the standard lightcone variables on the world-sheet
\be \xi^{\pm}\equiv{1\over 2}(\tau \pm \sigma),\quad \partial_{\pm}
\equiv \partial_{\tau}\pm\partial_{\si}\ee
and $\la.,.\ra$ denotes a non-degenerate invariant bilinear form on the 
Lie algebra $\r$ of $R$.
The second term in the WZNW action is commonly referred to as the
WZNW term and  it provide the action
with  the antisymmetric tensor part. It is well-known that
this antisymmetric tensor $B$   of the WZNW background
is not globally defined (for compact groups) because the
WZNW form $\o$ is a non-trivial cocycle in the third de Rham cohomology 
$H^3(R)$
of the         group manifold $R$. Inspite of this, the classical 
WZNW theory is well defined for the case of closed strings.
 The reason is simple: Consider an evolving loop
 which sweeps out a cylindrical world-sheet $g(\si,\tau)$
on the group manifold. 
The variational problem requires fixing
of the initial and  final position of the loop and slightly varying 
the position of
the cylinder  between: $r(\si,\tau)\to r(\si,\tau)+\delta r(\si,\tau);
~\delta r\vert_{initial,final}=0$.
 The antisymmetric tensor part of the variation of the
action can be thus written as 
\be \int (r+\delta r)^*B-\int r^*B = \oint dB=\oint 
\o.\ee
The integral $\oint$ is taken over the volume interpolating between 
the world-sheets $r$ and $r+
\delta r$ and * means the pull-back of the map. We conclude that the 
variation of the action does indeed depend
only on  the WZNW three-from $\o$ and not on a choice of its potential $B$. 
Note that the interpolating volume is given unambiguously because the 
variation of the action is infinitesimal. 

A well known additional  topological problem may occur if we wish
to define a path integral for the WZNW theory of closed strings \cite{Wit}:
Consider a set of fixed loops in $R$ and all world-sheets interpolating
among them. We wish to evaluate the WZNW action $S$ of every world-sheet $s$,
form an expression $\exp{iS}$ and sum up it over all interpolating world-sheets
of arbitratry topology. Suppose
we choose some reference interpolating world-sheet $s_{ref}$ and calculate 
its WZNW 
action
$S_{ref}$ for some choice of the potential $B$. The action $S$ of any other
world-sheet $s$ can be computed in the same way. It is tempting to conclude
that the difference $S-S_{ref}$ does not depend on the choice of the potential
$B$. Indeed, by using the same argument as in the variational problem, we
easily see that the difference of the integral of $B$ over the 
both world-sheets
is given solely in terms of the integral $\oint \o$ over the three-surface
which interpolates between the world-sheets\footnote{We shall alway assume that
the group manifold in question is simply connected. By Hurewicz isomorphism
and the fact the second homotopy group of any Lie group
vanishes we thus have that the second cohomology of the simply connected
group manifold
vanishes. This means that the interpolating three-surface always exists.}. 
But now the two world-sheets
 do not differ
only infinitesimaly! It therefore seems that the interpolating three-surface 
is 
not given unambiguously.  The way to get out of the trouble lies in comparing
the quantity $S-S_{ref}$ for two non-homotopical three-surfaces interpolating
between $s$ and $s_{ref}$. This difference is obviously given in terms of the
integral $\oint \o$ over a three-cycle obtained by taking the difference of 
(or the sum of oppositely oriented)  non-homotopical
three-surfaces interpolating between
$s$ and $s_{ref}$. Fortunately, the WZNW three-form $\o$ is an integer-valued
cocycle \cite{Wit} in $H_3(R)$ hence it is enough to normalize action $S$
properly in order to ensure that the quantities $S-S_{ref}$ differ by 
a term $2\pi k, k\in Z$ 
for any two interpolating three-surfaces. These $2\pi k$ terms  do not 
contribute to the path integral and, moreover, a dependence on the reference 
surface
$s_{ref}$ results only in an unobservable change of the total phase
of the path integral. We finish this little review 
 by concluding that the WZNW path integral is well
defined for the case of the interacting closed strings.

Consider now a $D$-branes configuration in the group target $R$. By this we
simply mean that there are two given submanifolds $D_i$ and $D_f$ of $R$ 
and  open  strings propagate on $R$ in such a way that their end-points
$i$ and $f$ stick on the $D$-branes $D_i$ and $D_f$, respectively.
We define the  WZNW theory for this $D$-branes configuration
 by choosing
 two-forms $\al_i$ and $\al_f$, living respectively on   $D_i$ and $D_f$
such that
\be d\al_{i(f)}= \o\vert_{D_i(D_f)} .\ee
In words: the exterior derivative of $\al_{i(f)}$ has to be equal to the
restriction of the WZNW three form $\o$ to the $D$-brane $D_{i(f)}$.

The construction of the $WZNW$ theory  based on the triplet $(\o,\al_i,\al_f)$
goes as follows: 
Pick up an open string $r(\si,\tau)$
with  the topology of an open strip. The variational 
problem requires fixing of the initial and the final positions
 of the string on the target. Consider now such a variation 
$\delta r(\si,\tau),
 \delta r(\si,\tau_{i,f})=0$.
The both original open strip and its variation form together a closed
strip (a `diadem'), whose edges lie on the opposite $D$-branes. We
can define the variation  $\delta S_{WZNW}$ of the WZNW term of the 
WZNW action 
by choosing an interpolating
surface $\Sigma_{i(f)}\ss D_{i(f)}$  between the edges of the original and 
the varied strip. This variation then reads
\be \delta S_{WZNW}=
\oint \o -\int_{\Sigma_i} \al_i - \int_{\Sigma_f} \al_f,\ee
where the $\oint \o$ is taken over the volume of the figure enclosed by 
$\Sigma_i$, $\Sigma_f$, the original strip and  its variation.
Note that this variation does not depend on the choice of the interpolating
surface $\Sigma_{i(f)}$ because $d\al=\o\vert_D$ and all infinitesimal
interpolating surfaces are mutually homotopic. Hence we conclude, that
the classical WZNW theory of open strings with end-points on the $D$-branes
is well defined in terms of the triplet  $(\o,\al_i,\al_f)$.

The reader may wish to have a more concrete idea of how to compute
the WZNW action of a single strip.
For a particular choice of the potential $B$ ($dB=\o$)
the combination $\al-B$ on the $D$-brane is a closed form, hence, at least
locally, it has a potential $A$ on $D$. The WZNW action $S$ for an open string
configuration $r(\si,\tau)$ which sweeps out a two-surface $s$ in the 
target $R$
and
     respects   the $D$-branes boundary conditions
can now be written as follows

\be 4\pi S(r)=\int \la \partial_+ r~r^{-1},
\partial_- r~r^{-1}\ra    +       \int_s B    +\int_{\delta s\cap D} A.
\ee
Upon a change of 
\be B\to B+d\lambda ,\ee 
 $A$ has to be replaced 
by 
\be A-\lambda\vert_D . \ee
We may intepret the $A$-term of the action (6) as if there were
equal and opposite charges on the end-points of the string
which feel the electromagnetic fields $A_i$ and $A_f$ on the $D$-branes.
This interpretation does not have an invariant meaning, however, because
of the `gauge invariance' (7) and (8). Moreover it holds
only locally. We stress that the {\it global}
invariant
description of the WZNW model for $D$-branes configuration is given
in terms of the triplet  $(\o,\al_i,\al_f)$. We remark that in general
there is no natural {\it closed} two-form living on the $D$-branes. This
is true only in the case if the restriction of the WZNW three-form $\o$
on the $D$-brane vanishes.  Note also that if   the $D$-brane is as many 
dimensional as  the whole group 
target  
$R$ is, then the form $\al$ is nothing but some concrete choice of the 
potential
$B$ which, however, may be different for the different end-points
of the string. 

At the presence of the $D$-branes and open strings, the discussion of the 
string path integral is more involved as before.
The  group manifold will be  always taken as 
 simply connected and, for a while,
we consider the case where also the $D$-branes are connected and 
simply connected.
 Now draw a general string
diagram respecting the $D$-branes configurations. It is an interpolating 
world-sheet between a set of fixed open segments with end-points
located on the $D$-branes
and a fixed  set of loops on the target $R$. Much as before, we can choose
some reference interpolating world-sheet $s_{ref}$ and calculate
its  WZNW part of the 
action $S_{ref}$ for some choice of $B$ and $A$ according to the
formula (6). Now we can take any other interpolating world-sheet $s$ and 
calculate
its  action $S$ in the same way. As in the
case of the variational principle, the quantity $S-S_{ref}$ does not depend
on the particular choice of $B$ and $A$ but only on the invariant
globally defined triplet $(\o,\al_i,\al_f)$. The reason for this is the 
following: the union of the intersections $(\partial s_{ref}\cap D_{i(f)})
\cup (\partial s\cap D_{i(f)})$ is a contractible cycle in $D_{i(f)}$, 
hence it is a boundary of
some two-surface $\Sigma_{i(f)}$. Now the union $s\cup s_{ref}\cup\Sigma_i\cup
\Sigma_f$ is a two-boundary of some interpolating three-surface
in the group manifold, because the second 
cohomology of the group manifold vanishes by assumption. Then 
the antisymmetric tensor (the WZNW term) part  of $S-S_{ref}$
is defined by (5) where $\oint$ is taken over the interpolating
three-surface.

There occurs the same
problem as for the closed strings, namely, the interpolating three-surfaces
between $s$ and $s_{ref}$ do not have to be homotopically
equivalent. This means that the quantity $S-S_{ref}$ may depend on the
homotopy of the chosen interpolating three-surface. But if the ambiguity
in $S-S_{ref}$ is only of the form $2\pi k, k\in Z$ then the 
term $\exp{i(S-S_{ref})}$  is unambiguous and the path integral
is well defined. 

It is not difficult to find a cohomological formulation
of  the condition of the integer-valued
ambiguity. All what we need is the notion of the
relative singular  homology $H_*(R,D_i\cup D_f)$ 
of the manifold $R$ with respect to its submanifolds
$D_i$ and $D_f$ (with real coefficients).
 The relative  chains are the elements of the vector
space of the standard
chains in $R$  factorized by its subspace of all chains lying in $D_i\cup D_f$.
The operation of taking the boundary is the standard one. The corresponding
homology is the relative singular homology  $H_*(R,D_i\cup D_f)$.
The triplet $(\o,\al_i,\al_f)$ can act on a relative cycle $\gamma$  by 
the following prescription 
 \be \la (\o,\al_i,\al_f),\gamma\ra\equiv \int_{\gamma}\o
-\int_{D_i\cap\partial\gamma}\al_i -\int_{D_f\cap\partial\gamma}\al_f.\ee
If the cycle $\gamma$ is itself a boundary then the pairing vanishes
because $\o$ is closed. Hence 
our triplet $(\o,\al_1,\al_f)$ is an element (cocycle)
of the relative singular cohomology  $H^*(R,D_i\cup D_f)$ 
 because it vanishes
on the boundary of any relative chain. 

Now we may conclude that if the cocycle $(\o,\al_1,\al_2)$ 
is integer-valued\footnote{The precise statement is as follows: The
cocycle
$(\o,\al_1,\al_2)$  is integer-valued, if it lies 
in the image of the natural map from
the singular cohomology with integer coefficients to the singular
cohomology with real coefficients.} the WZNW path integral is well-defined.
Indeed, if we choose two non-homotopical three-surfaces interpolating
between the world-sheets $s$ and $s_{ref}$ their oriented sum is
a closed cycle in the relative singular homology and its pairing (9) 
with the triplet $(\o,\al_1,\al_2)$  is   integer-valued.

It turns out that we can extend our discussion to the case of connected
but not necessarily simply connected $D$-branes. 
The main problem to be addressed is the fact that now the
union of the intersections
 $(\partial s_{ref}\cap D_{i(f)})
\cup (\partial s\cap D_{i(f)})$ is not necessarily a contractible cycle in 
$D_{i(f)}$
(which means that $s\cup s_{ref}$ is a relative two-cycle
but not a relative two-boundary).
Thus the two-surface $\Sigma_{i(f)}$ does not have to exist and we cannot
in general use the formula (5) in order to determine $\exp{i(S-S_{ref})}$.
It may seem that we may take some reference world-sheet 
for each homotopy class of the one-chain $\partial s\cap D_{i(f)}$ and assign
it an arbitrary reference phase. But there is still a consistency condition
that under summing of the relative two-cycles (unions of $s$ and $s_{ref}$)
the phases $\exp{iS}$ should be  additive!

 Recall that  we can
unambiguously assign the $\exp{iS}$ to every relative two-boundary
in such a way that this mapping is homomorphism $f$ from the group $B$ of 
relative two-boundaries (with integer coefficients) into the group $U$
of complex units (phases). The consistency condition means
that there should exist an extension $\ti f: Z\to U$ 
of this 
homomorphism defined on the group $Z$ of all relative two-cycles. We now 
prove that such
an extension always exists because $U$ is the divisible group (this means
that the equation $nx=a, a\in U, n\in {\bf N}$ has always a solution $x\in U$).

Consider the group $H_f=Z+U/\{b-f(b),b\in B\}$. We have an exact sequence
\be 0\to U\to H_f \to H\to 0,\ee
where $H\equiv H_2(R, D_i\cup D_f)=Z/B$ and all homomorphisms are naturally
defined. Suppose now that we do have an extension $\ti f:Z\to U$ of the map
$f:B\to U$. Such an  extension enables us to write 
\be H_f= H+U.\ee
In words: $H_f$ is a direct sum of $H$ and $U$. Indeed, for 
$z+c, z\in Z,c\in U$
we have
\be z+c = (z- \ti f(z))  +(0 +c+\ti f(z)).\ee
Evidently, the first term on the right hand side is from $Z$ and the second
from $U$. The decomposition (12) is consistent with the factorization
by $\{b-f(b), b\in B\}$ because $\ti f$ is the homomorphism.
The converse is also true: if we can write $H_f$ as the direct sum
$H+U$ then there exists an extension $\ti f:Z\to U$ which is a homomorphism.
Indeed, consider $z\in Z$ and embed it naturally into $H_f$ i.e.
$z\to z+0\in H_f$. $z+0$ can be decomposed as $y+g, y\in H, g\in U$
by assumption, hence we obtain a natural homomorphism from $Z$ into $U$:
$z\to g$. This homomorphism is the extension of $f$ which we look for.

Summarizing, if we prove that $H_f$ is the direct sum of $H$ and $U$,
we are guaranteed that the extension $\ti f:Z\to U$ always exists. But it is
easy to prove this, by using the well-known result from the homological
algebra that every extension of an (Abelian) group G
by a divisible group $X$ is necessarily
the direct sum of $G$ and $X$. In our case, we know from 
the exact sequence (10) that $H_f$ is the extension of $H$ by $U$. Therefore
$H_f=H+U$, what was to be proved.

\noindent {\it Notes}:

\noindent 1.  
We have a certain freedom in writing $H_f$ as a direct sum of $H$ and $U$
which is 
 described by the group of homomorphisms $Hom(H,U)$. The easiest way to see
it is by noting that if we have an extension $\ti f:Z\to U$ it can be modified
by adding to it any homomorphism which vanishes on $b\in B$. Any 
such homomorphism is obviously from $Hom(H,U)$. The modified $\ti f$ then
gives another partition of $H_f$ into the direct sum of $H$ and $U$. 

\noindent 2. It may be instructive to relate the group $H$ of the relative
two-cycles with the fundamental  groups $\pi_1$ of the $D$-branes.
We have a natural exact sequence
\be 0=H_2(R)\to H_2(R,D_i\cup D_f)\to H_1(D_i)+H_1( D_f)\to 0=H_1(R).\ee
Hence
\be H=H_1(D_i) +H_1(D_f)\ee
and
\be H_1(D_{i(f)})=\pi_1(D_{i(f)})/[\pi_1(D_{i(f)}),\pi_1(D_{i(f)})].\ee
The last equality is the Hurewicz isomorphism which holds due to the assumption
that the $D$-branes are connected.

\section{PL symmetries of  WZNW models}

For the description of the PL $T$-duality, we need the
crucial concept
of the Drinfeld double,  which is simply  a  Lie group $D$ such that
its Lie algebra $\cd$ (viewed as a vector space) 
 can be decomposed as the direct sum  of two  subalgebras, $\cg$ and $\cgt$, 
maximally isotropic with 
respect to a non-degenerate invariant bilinear form on $\cd$ \cite{D}.
It is often convenient to identify the dual linear space to $\cg$ ($\cgt$)
with $\cgt$ ($\cg$) via this bilinear form.

From the space-time point of view, we have identified 
 the targets of the mutually 
dual $\sigma$-models
with the cosets $D/G$ and $D/\ti G$ \cite{KS6}. 
Here $D$ denotes the Drinfeld double,
and $G$ and $\ti G$ two its mutually dual isotropic subgroups.  In the special
case when the decomposition $D=\ti G G= G\ti G$ holds globally, the 
corresponding
cosets turn out to be the group manifolds $\ti G$ and $G$, respectively 
\cite{KS2}.

The actions of mutually dual $\si$-models 
are encoded in a choice of an $n$-dimensional
linear subspace ${\cal R}$ of the $2n$-dimensional
Lie algebra $\cd$ of the double $D$ which is transversal to both $\cg$ and 
$\cgt$. The $\si$-model actions 
on the targets $D/G$ and $D/\ti G$
have a similar structure; indeed, on $D/G$ we have \cite{KS6}
\be S={1\over 2}I(f)-{1\over 4\pi}\int d\xi^+ d\xi^- \la 
\partial_+ f~f^{-1},R_-^a\ra (M_-^{-1})_{ab}\la f^{-1}\partial_- f,T^b\ra,\ee
where $f\in D$ is some local section of the $D/G$ fibration which
parametrizes the points of the coset. Recall \cite{KS6} that
\be M_{\pm}^{ab}\equiv \la T^a ,f^{-1}R_{\pm}^b f\ra\ee
and $R_-^a$ ($R_+^a$) are vectors of an orthonormal basis of ${\cal R}$ 
(${\cal R}^{\perp}$):
 \be \la R_{\pm}^a,R_{\pm}^b\ra= \pm\delta^{ab},\qquad \la R_+^a,R_-^b\ra=0.\ee

 The action of the dual $\sigma$-model on the coset $D/\ti G$ has 
the same form; just the generators $T^a$ of $\cg$ are replaced by the 
generators 
$\ti T_a$ of $\cgt$ and $f$ will parametrize $D/\ti G$ instead of $D/G$.

We have referred to the $\si$-models of the form  (16) as
those  having a PL symmetry \cite{KS6}.  There is an important feature
of such models, namely,  their  field equations
can be written as the zero curvature condition valued in the algebra
$\cg$. Indeed, 
\be d\lambda-\lambda^2=0,\ee
where
\be \lambda =\lambda_+ d\xi^+ +\lambda_- d\xi^-\ee
and
\be \lambda_{\pm}=-\la \partial_{\pm} f~f^{-1},R_{\mp}^a\ra
 (M_{\mp}^{-1})_{ab}T^b.\ee

So far we have been reviewing the results of \cite{KS6}; now a new
observation comes: If the subspace ${\cal R}$ is itself a Lie algebra
of a compact subgroup $R$ of the double $D$ then 
the model (16) is essentially the WZNW model on the target $R$ for the both 
choices $D/G$ and $D/\ti G$!  
The argument goes in two steps: 

\noindent 1.  ${\cal R}$ 
can be transported by the right action to the 
tangent space of every point of the double.
Because ${\cal R}$ is the subalgebra, the distribution of the planes ${\cal R}$
in the tangent bundle of the double is integrable and it foliates
the double into fibration with fibres $R$ and basis $R\backslash D$.
Since ${\cal R}$ is transversal
to the both $\cg$ and $\cgt$ (which means that it intersects $\cg$ and
$\cgt$ only in $O$) , any fiber of the $R$ fibration
either intersects the fiber $G$ (or $\ti G$) in some finite subgroup 
$R\cap G$ of 
$R$
or does not intersect it at all. The latter cannot be true, however,
if the group $R$ is compact. Indeed, $R$ acts on $D/G$ by the left action. 
The $R$ orbit of the element of $D/G$ which has the unit
element of $D$ on its fiber is open. Since $R$ is compact this orbit
must be also closed which for connected doubles
imply that this orbit is the whole $D/G$. In other words,
there always exists an intersection of $R$ and $G$.
The argument for $D/\ti G$ is the same.

If the finite subgroups $R\cap G$ and $R\cap \ti G$  have  only one element
 for 
both fibers $G$ and $\ti G$, respectively, it si not dificult to see 
that the both  cosets $D/G$ and $D/\ti G$ can be globally 
identified with $R$. In general, the cosets $D/G$ and $D/\ti G$ can be 
identified with the discrete cosets $R/R\cap G$ and $R/R\cap  \ti G$,
respectively.

\noindent  2. For simplicity, consider only the case when $R$ can be directly
identified with $D/G$ and $D/\ti G$. In this case, we can choose 
the field $f(\si,\tau)$ in (16) to have values in $R$. Note that 
we can choose the basis $R_-^a$ dependent on $f$ in such a way that 
the combinations $f^{-1}R_-^a f$ are $f$ independent. Then we can choose
the basis $T^a$ in such a way that $M_-(f)$ is the identity matrix.
We have 
\be \la \partial_+ f ~f^{-1}, R_-^a\ra =\la f^{-1} \partial_+ f, f^{-1}R_-^a
f\ra\equiv (f^{-1} \partial_+ f)^a\ee
and
\be   \la f^{-1}\partial_- f , T^a\ra=
\la f^{-1}\partial_- f,f^{-1}R_-^c f\ra M_-^{ca}=(f^{-1}\partial_- f)^a,\ee
because $M_-$ is the identity matrix. Putting (16),(22) and (23) together,
we obtain
\be S={1\over 2}I(f)-{1\over 4\pi}\int d\xi^+ d\xi^-
(\partial_+ f ~f^{-1})^a \delta_{ab}(\partial_- f ~f^{-1})^b=-{1\over 2}
I(f^{-1}
).\ee
We conclude, that the mutually dual $\si$-models on the cosets $D/G$ and 
$D/\ti 
G$ are the same, being equal to the WZNW model on $R$. In general,
$D/G$ ($D/\ti G$) model is WZNW model on the target $R/R\cap G$
($R/R\cap \ti G$).

\noindent {\it Notes}: 

\noindent 1. The fact that the both models $D/G$ and $D/\ti G$ may be
identical does not mean at all that the duality transformation is trivial.
In fact, the PL $T$-duality always implies an existence of 
a  non-trivial non-local 
transformation on the phase space of the $WZNW$ model. We shall explicitly
describe this transformation in the next section.

\noindent 2. It often happens (cf. section 4) that a compact group
$R$ can be embedded in many inequivalent ways into various Drinfeld
doubles in such a way that the both cosets $D/G$ and $D/\ti G$
can be identified with $R$. In this case we have the abundance
of the Poisson-Lie symmetries of the same WZNW model on the group manifold
$R$, each of them corresponding to the double into which $R$ is embedded.

\section{$D$-branes in WZNW models}
\subsection{General discussion}

For the further discussion of the $D$-branes, it is convenient to 
recall \cite{KS6} the common `roof' of the both models described by (16).
They can be derived form the first order Hamiltonian action for 
field configurations $l(\si,\tau)\in D$:
$$ S[l(\tau,\si)]= $$
\be ={1\over 8\pi}\int  \biggl\{\la \pas l~l^{-1},\pat l~l^{-1}\ra+
{1\over 6}d^{-1}\la dl~l^{-1},[dl~l^{-1},
dl~l^{-1}]\ra -\la \pas l l^{-1},A\pas l l^{-1}\ra \biggl\}.\ee
Here $A$ is a linear idempotent self-adjoint map from the Lie algebra
$\cd$ of the double into itself. It has two equally degenerated
eigenvalues $+1$ and $-1$, and the corresponding eigenspaces are just
${\cal R}^{\perp}$ and ${\cal R}$ respectively.

 As it stands, the action (25)
is well defined only for the periodic functions of $\sigma$ because of the
WZNW term. This restriction corresponds to the case of closed strings 
\cite{KS6} . The $\si$-model actions (16) are obtained from 
the duality invariant first order action (25) as follows:
 Consider  the  right coset
$D/G$ and parametrize it by the elements $f$ of $D$ \footnote{If there
exists no global section of this fibration, we can choose several
local sections covering the whole base space $D/G$.}.
With this parametrization of $D/G$ we may parametrize the surface
$l(\tau,\sigma)$ in the double as follows
\be l(\tau,\sigma)= f(\tau,\sigma)g(\tau,\sigma),\quad  g\in G.\ee
The action $S$ then becomes 
$$ S(f,\Lambda\equiv \pas g g^{-1})={1\over 2}I(f) -{1\over 2\pi}
\int d\xi^+ d\xi^- \biggl\{\big\la \Lambda -
{1\over 2} f^{-1}\d- f , \Lambda -{1\over 2}f^{-1}\d- f \big\ra$$
\be +\la f\Lambda f^{-1} +\pas f f^{-1}, R_-^a\ra\la R_-^a , f\Lambda f^{-1}
+\pas f f^{-1}\ra\biggl\}.\ee 
Now it is easy to eliminate $\Lambda$ from the action (27) and finish with
the $\si$-model action (16). In the case of the coset $D/\ti G$, the procedure
is exactly analoguous.

Consider  the case of open strings for a generic double $D$ with vanishing
second cohomology. In our
previous paper on the subject \cite{KS4}, we have studied only
the perfect doubles (cf. footnote 3) nevertheless we can easily generalize
the construction.

  Let $F$ be a simply connected subgroup of the double $D$ whose Lie
algebra ${\cal F}$ is isotropic with respect to the bilinear form on $\cd$.
This subgroup, as a manifold, can be shifted by the right action of some
element $d\in D$ (note that all non-equivalent shifts are parametrized by 
the coset $F\backslash D$). We declare that the manifolds $F\hookrightarrow
D$ and $Fd\hookrightarrow D$ are  $D$-branes in the double $D$.
Consider now oriented open strings in $D$ with the initial end-points
on $F$ and the final end-points on $Fd$. Their dynamics in the bulk is
 governed by the action (25) which contains the WZNW term. As we have learnt
in the previous section such an action  is well-defined provided
we choose some two-forms on the $D$-branes such that the exterior derivative
of them is equal to the restriction of the $WZNW$ three-form on the 
$D$-branes. In our present case, this restriction of the WZNW form
vanishes in either of our $D$-branes because 
$F$ and $Fd$  are the isotropic surfaces in $D$.
Thus we have to choose some closed two forms on $F$ and $Fd$; we choose them
to vanish identically. We summarize that our open string dynamics
is fully defined by the action (25), the $D$-branes boundary conditions
and the vanishing two-forms on the $D$-branes.

Much as in the closed string case, we can derive the open string $\si$-model 
dynamics on the cosets $D/G$ and $D/\ti G$ from (25) and the $D$-branes
data on the double; for concreteness let us consider the coset $D/G$:

As we have learnt in section 2, the WZNW model for open strings
is fully defined if we manage to compute the WZNW action
of the `diadem'. Recall that the  diadem  is composed of two evolving
open string world-sheets which are glued together at some initial
and final times. The edges of the  diadem , swept by the end-points
of the open strings, lie in their corresponding $D$-branes. 

Consider now the diadem  in the double. We can choose some two-surface
$\Sigma$ ($\Sigma_d$) in the $D$-brane $F$ ($Fd$) whose boundary is just the
 edge
of the diadem lying in $F$ ($Fd$). The diadem together with the surfaces
$\Sigma$ and $\Sigma_d$ form a boundary of some three-dimensional domain 
$\gamma$.
We may write the action $S$ of the model (25) as
\be S=S_0+S_{WZNW},\ee
where $S_{WZNW}$ contains solely the term with the WZNW three-form $c$ on $D$.
 Hence, the action of the diadem can be written as\footnote{Note that we 
have included the factor $1/6$ from (25) in the definition of $c$.}
\be S=S_0 + {1\over 8\pi}\int_{\gamma}c.\ee

 Again, 
consider  the  parametrization of 
$D/G$  by the elements $f$ of $D$.
 A surface
$l(\tau,\sigma)$ in the double (respecting the $D$-branes boundary conditions),
 can be written as 
\be l(\tau,\sigma)= f(\tau,\sigma)g(\tau,\sigma),\quad  g\in G.\ee
The
decomposition (30) induces two maps from $D$ into $D$: $f(l)=f$
and $g(l)=g$. Consider now the Polyakov-Wiegmann (PW) formula \cite{PW}
\be (fg)^*c=f^*c +g^*c -d\la f^*(l^{-1}dl) \stackrel {\w}{,}g^*(dl ~l^{-1})
\ra,\ee
where, as usual,  $*$ denotes the pull-back of the forms
under the mappings to the group manifold $D$.
By using the PW formula, we can rewrite (29) as
\be S=S_0(fg) +{1\over 8\pi}\int_{\gamma} f^*c -{1\over 8\pi}\int_{diad\cup 
\Sigma
\cup \Sigma_d}\la f^*(l^{-1}dl)\stackrel {\w}{,} g^*(dl~l^{-1})\ra.\ee
Note that $g^*c$ vanishes because of the isotropy of $G$. The action $S$ now
becomes

$$ S(f,\Lambda\equiv \pas g g^{-1})={1\over 2\pi}\int_{diad}\biggl\{
{1\over 4}\la  \partial_+ f~f^{-1},\partial_- f~f^{-1}\ra$$
$$ -\big\la \Lambda -
{1\over 2} f^{-1}\d- f , \Lambda -{1\over 2}f^{-1}\d- f \big\ra
 +\la f\Lambda f^{-1} +\pas f f^{-1}, R_-^a\ra\la R_-^a , f\Lambda f^{-1}
+\pas f f^{-1}\ra\biggl\}$$
\be +{1\over 8\pi}\int_{\gamma}f^*c
-{1\over 8\pi}\int_{\Sigma\cup \Sigma_d}\la f^*(l^{-1} dl)\stackrel {\w}{,} 
g^*(dl~l^{-1}
)\ra.\ee 
Of course, this is  a similar  expression as before (cf. (27)).  However,
the field $f$ respects different boundary conditions. A configuration
$f$ is an open string configuration; its end-points stick on $D$-branes
$D_i$ and $D_f$ in $D/G$ which are obviously obtained just by projecting the 
$D$-branes
$F$ and $Fd$ from the double D into the basis $D/G$ parametrized by the 
section $f$.

 Now  we have to realize that upon varying $\Lambda$
 the last term in (33) vanishes! This follows from the 
isotropy of $F$, $Fd$ and $G$. Indeed, if we have $fg\in F$ ($fg\in Fd$)
and  vary $g\to g\delta g$
at fixed $f$ in such a way that  $fg\delta g\in F$ ($fg\delta g \in Fd$),
we observe that the last term in (33) does not change\footnote{It is easy
to see that $\delta g\in F\cap G$ ($\delta g\in F\cap dGd^{-1}$).}.
Hence we can  eliminate the field $\Lambda$ from (33) 
in the same way as from (27). The result is

$$ S={1\over 8\pi}\int_{diad}
\biggl\{\la  \partial_+ f~f^{-1},\partial_- f~f^{-1}\ra -2 \la 
\partial_+ f~f^{-1},R_-^a\ra (M_-^{-1})_{ab}\la f^{-1}\partial_- f,
T^b\ra\biggl\}$$
$$ +{1\over 8\pi}\int_{\gamma}f^*c
-{1\over 8\pi}\int_{\Sigma\cup \Sigma_d}\la f^*(l^{-1}dl)\stackrel {\w}{,} 
g^*(dl~l^{-1})
\ra.$$

 Consider again the special situation in which the subspace ${\cal R}\equiv
{\rm Span} R_-$
is the Lie algebra of the compact group $R$, moreover, $R$ can be directly
identified with $D/G$ and $D/\ti G$. Recall, that upon transporting
${\cal R}$ by the right action everywhere onto the double, we get the 
fibration of $D$ with the fibers $R$ and the basis $R\backslash D$.
With some abuse of the notation, the fiber crossing the unit element
of the double we shall also denote as $R$.
We choose the parametrization of the double as follows
\be l=rg,\quad r\in R,\quad g\in G.\ee
This parametrization holds for every element $l$ of the double
and is unique by the assumption. Note that the restriction of the
 WZNW three-form $c$  gives just the WZNW three-form $c_R$ on $R$.

It is easy to see that the $D$-branes $D_i$ and $D_f$ in $R$, being the 
projections
of $F$ and $Fd$ to $R$, can be identified with the cosets
$F/F\cap G$ and $F/F\cap dGd^{-1}$ respectively. On the other hand we have 
just seen (cf. footnote 9)  that the variation $\delta g\in F\cap G$ 
($\delta g
\in F\cap dGd^{-1}$) leaves intact the two-form 
$\omega \equiv (1/8\pi)\la r^*(l^{-1}dl)\stackrel {\w}{,} g^*(dl~l^{-1}
)\ra$ on $F$ ($Fd$). This  means that this two-form is a pull-back of
some two-form $\al_i$ ($\al_f$) from the $D$-brane $D_i$ ($D_f$).
Of course, the notation is not accidental; the two-forms $\al_{i(f)}$ are 
precisely those appearing in (5). 

It is not difficult  to find an explicit
expression for $\al_{i(f)}$. For this, consider a map $k_i$ ($k_f$)
from $D_i$  ($D_f$) into $G$ such that
\be rk_{i(f)}(r) \in F(Fd), \quad r\in D_{i(f)}.\ee
In general, the mapping $k_i$ ($k_f$) is not
 defined  unambiguously but it locally always exists since $D_i$ ($D_f$)
is just  the projection of $F$ ($Fd$) on $R$.
Because two-form $\omega$ on $F$ ($Fd$) is invariant
under the variations from $F\cap G$ ($F\cap dGd^{-1}$) we can  
locally\footnote{The two-form $\al_{i(f)}$ is  defined {\it globally}
on $D_{i(f)}$ only the explicit expression for it in terms of $k_{i(f)}$ 
may, in general, be written only locally.} write 
\be   \la r^*(l^{-1}dl)\stackrel {\w}{,} g^*(dl~l^{-1})\ra\vert_{F(Fd)}=
r^*\la dr ~r^{-1}\stackrel {\w}{,} k_{i(f)}(r)^{-1}dk_{i(f)}(r)\ra.\ee
 In other
words, (36) is true independently of the choice of the map $k_{i(f)}$. 

Thus in our special situation, the action of the diadem can be written
as 
$$ S=-{1\over 8\pi}\int_{diad}
\la  \partial_+ r~r^{-1},\partial_- r~r^{-1}\ra +
{1\over 8\pi}\int_{r(\gamma)}c_R $$
\be -{1\over 8\pi}\int_{D_i}\la dr ~r^{-1}
\stackrel {\w}{,} k_i(r)^{-1}dk_i(r)\ra 
-{1\over 8\pi}\int_{D_f}\la dr ~r^{-1}
\stackrel {\w}{,} k_f(r)^{-1}dk_i(r)\ra.\ee

 We can read   $\al_{i(f)}$ off directly from (37):
\be \al_{i(f)}= {1\over 8\pi}\la dr ~r^{-1}
\stackrel {\w}{,} k_{i(f)}(r)^{-1}dk_{i(f)}(r)\ra.\ee
It remains 
to prove that 
\be d\al_{i(f)}={1\over 8\pi} c_R\vert_{D_{i(f)}}.\ee
This is easy: take the PW formula (31) and restrict all forms
in it on the $D$-brane $F$ ($Fd$) in the double. Then the form $c$ vanishes
by the isotropy of $F$ ($Fd$). Hence
\be r^*c\vert_{F(Fd)}=d\la r^*(l^{-1}dl)\stackrel {\w}{,} g^*(dl~l^{-1})
\ra\vert_{F(Fd)}=r^*d\la dr ~r^{-1}\stackrel {\w}{,} 
k_{i(f)}(r)^{-1}dk_{i(f)}(r)\ra,\ee
where the last equality follows from (36).  Thus,
 upon removing the pull-back map
$r^*$, we conclude that
\be {1\over 8\pi}c_R\vert_{D_{i(f)}}={1\over 8\pi}d\la r^{-1} dr
\stackrel {\w}{,} 
dk_{i(f)}(r)k_{i(f)}(r)^{-1}\ra=
d\al_{i(f)}.\ee

\noindent {\it Remarks}: 

\noindent 1. The model (37) has the `wrong' sign in front of its 
first
term. Upon the change of variables $r\to r^{-1}$ it gives the
standard WZNW model on the group manifold $R$ (cf. (1)). 
The $D$-branes $D_{i(f)}$
and the two
forms $\al_{i(f)}$ on them have to be transformed correspondingly.

\noindent 2. The geometry of the dual $D$-branes in $D/\ti G$ is obtained
in the same way as in the case $D/G$; it is enough to replace
everywhere $G$ by $\ti G$.

\noindent 3. We should mention that the Kiritsis-Obers duality \cite{KO} fits
in our formalism. The double is the direct product of a compact
group $R$ with itself and the invariant bilinear form 
in the direct sum of the Lie algebras $\r+\r$ is the difference
between the Killing-Cartan forms on each algebra. Hence, the diagonal
embedding of $R$ in $R\times R$ is isotropic. So it is the 
embedding in which second copy of $\r$ is twisted by some outer automorphism.
The resulting duality is a $D$-branes $D$-branes duality, i.e. the 
$D$-branes have never the dimension of the group manifold.

\subsection{The classical solvability}

We wish to find the complete solution of the field equations
of the model (25) submitted to the $D$-branes boundary conditions.
It is not difficult to do that. The bulk equations following from (25) read
\be \la \partial_{\pm}l ~l^{-1}, {\cal R}_{\mp}\ra=0.\ee
  We already know that after integrating away $g$ from the
decomposition (34)  we get the WZNW model
on $R$, hence, the solution $l$ of (25) must look
like
\be l(\si,\tau)=r_-(\xi^-)r_+(\xi^+)g(\xi^+).\ee
The first two multiplicative terms on the right-hand-side follow from
the known bulk solution of the WZNW model on $R$ and the fact that
$g$ is only a function of $\xi^+$ follows from Eqs. (21).

Putting
\be h(\xi^+)\equiv r_+(\xi^+)g(\xi^+)\ee
and inserting $l=r_-(\xi^-)h(\xi^+)$ into (37),
we obtain
\be \partial_+h ~h^{-1}\in {\cal R}_+\equiv {\cal R}^{\perp}.\ee
Here we have used the fact that ${\cal R}_-(\equiv {\cal R})$ is 
the Lie algebra of $R$. We conclude that every bulk solution of
(25) look like
\be l=r_-(\xi^-)h(\xi^+), \qquad \partial_+h~h^{-1}\in \r_+.\ee
It is important to note that $\r_+\equiv\r^{\perp}$ does not have to be a 
Lie subalgebra
of $\cd$; in general it is just a linear subspace of $\cd$.

Now we can take into account the effect of the boundary conditions.
Recall that the initial point of the open string ($\si=0$) should
stick on the $D$-brane $F$ in the double and the final point ($\si=\pi$)
on the $D$-brane $Fd$; $d$ is a fixed element of the double $D$.
These two conditions can be rewritten as follows
\be r_-(\tau)h(\tau)=f_i(\tau),\qquad r_-(\tau-\pi)h(\tau)=f_f(\tau)d,\ee
where $f_i$ and $f_f$ are some functions with values in the group $F$.
It follows that 
\be h^{-1}(\tau-\pi)h(\tau)=f_i^{-1}(\tau-\pi)f_f(\tau)d.\ee
By differentiating Eq. (48) with respect to $\tau$ we obtain
$$ -dh(\tau-\pi)h^{-1}(\tau-\pi)+dh(\tau)h^{-1}(\tau)=$$
\be =h(\tau-\pi)[-f_i^{-1}(\tau-\pi)df_i(\tau-\pi) +f_i^{-1}(\tau-\pi)
df_f(\tau)f_f^{-1}(\tau)f_i(\tau-\pi)]h^{-1}(\tau-\pi).\ee
Now we can bracket (49) with $\r$ which gives
\be df_f(\tau)f_f^{-1}(\tau)-df_i(\tau-\pi)f_i^{-1}(\tau-\pi)=0.\ee
For deriving (50), we have used Eq. (47) and the fact
that the Lie algebra $\cf$ of
$F$ is  transversal to $\r^{\perp}$.

By inserting (50) back in (49) we get a very important relation
\be dh(\tau+\pi)h^{-1}(\tau+\pi)=dh(\tau)h^{-1}(\tau).\ee
It expresses the periodicity of the $\r^{\perp}$-valued `connection'
$dh~h^{-1}$. The monodromy of this `connection is also constrained; indeed,
from (50) and (48) we conclude that
\be h^{-1}(\tau-\pi)h(\tau)=fd,\ee
where $f$ is some constant element of $F$. In words: the monodromy  
$h^{-1}(\tau-\pi)h(\tau)$
is an element of the double $D$ which is equivalent to $d$ in the
sense of the coset $F\backslash D$.

\noindent {\it Summary}: The space of the solutions of the
field equations (42) submitted to the $D$-branes boundary conditions (47)
is given by an arbitrary element $p$ of the double $D$ and a
 periodic field $\rho(\xi^+)(\equiv dh~h^{-1}(\xi^+))$
with values in the subspace $\r^{\perp}$ of $\cd$ and with  the monodromy

\be  P\exp{\int_{\tau -\pi}^{\tau}d\tau' \rho(\tau')}\equiv 
 h^{-1}(\tau-\pi)h(\tau)\ee
equivalent to $d$ in the sense of the coset $F\backslash D$. Of course,
$P$ in (53) means the ordered exponent.
The full solution $l(\si,\tau)$ is then reconstructed as follows:
take $\rho(\xi^+)$ and $p\in D$ and construct
\be h(\xi^+)=P\exp{\{\int_{\xi_0^+}^{\xi^+}d\xi^{+'} \rho(\xi^{+'})\}}\times
p.\ee
Obviously, the choice of $\xi^+_0$ is irrelevant and can be compensated
by the corresponding change of $p$. Now we can reconstruct $r_-(\xi^-)$
by decomposing  $h(\xi^-)$ as
\be h(\xi^-)=r_-^{-1}(\xi^-)f(\xi^-), \qquad r\in R,\quad f\in F. \ee
This decomposition is unique, because $R$ can be globally identified
with $D/F$. Finally
\be l(\xi^+,\xi^-)=r_-(\xi^-)h(\xi^+).\ee

It remains to recover from the solution (56) on the double the solutions
of the $\si$-models on the cosets $D/G$ and $D/\ti G$. Recall that
the both  $\si$-models are the WZNW models on the group manifold $R$.
In the $D/G$ case we have to decompose $h$ as
\be h(\xi^+)=r_+(\xi^+)g(\xi^+), \qquad r\in R, \quad g\in G,\ee
while in the $D/\ti G$ case as
\be h(\xi^+)=\ti r_+(\xi^+)\ti g(\xi^+), \qquad \ti r\in R, 
\quad \ti g\in \ti G.\ee
Because $r_+\neq 
 \ti r_+$ we indeed obtain a nontrivial map from
the phase space of the WZNW model with one set of the $D$-branes
boundary conditions into the phase space of the same WZNW model
but with the dual $D$-branes boundary conditions. The both phase
spaces can be identified with the set of all solutions $l(\xi^+,\xi^-)$
on the double. The system (25) is already written in the Hamiltonian
form, hence the mapping between the phase spaces is a canonical 
transformation.

\noindent {\it Note}:  It is interesting that the both left movers 
$r_+(\xi^+)$
and right movers $r_-(\xi^-)$ are obtained from the master function
$h(\xi^+)$ in a very similar way. Recall that
\be  h(\xi^+)= r_+(\xi^+) g(\xi^+), \qquad  h(\xi^-)= r_-^{-1}(\xi^-) 
f(\xi^-), \quad g\in G,~f\in F.\ee
In particular, if $G=F$ then the left and the right movers of the
$R$ WZNW model are given
by the same function , i.e.
\be r(\si,\tau)=r_-(\xi^-)r_-^{-1}(\xi^+).\ee
This means that the initial point $\si=0$ of the string sits at the origin
of the group $R$ for all times. Indeed, the corresponding $D$-brane
is just the group origin, being the projection of $F=G$ along $G$.

\subsection{Interacting $D$-brane diagrams}

Given a $D$-brane configuration on the target $R$ we can in principle
compute                                          the WZNW path integral
over all topologically non-trivial world-sheets interpolating between
a set of fixed
open string 
segments with  end-points sitting on the  $D$-branes and a set of fixed
loops in the target $R$.
We postpone the evaluation of 
some of such diagrams (like  the open string 
propagator) to a forth-coming publication,
here we just discuss whether there are some topological obstructions
in doing that possibly
coming from the WZNW term $\o$ and the two-forms $\al_i$ and $\al_f$ defined
on the $D_i$ and $D_f$ by (38). We have learned in the section 2 that
the WZNW path integral is well-defined if the triplet $(\o,\al_i,\al_f)$
is an integer-valued cocycle in the relative singular cohomology
of the group manifold $R$ with respect to its submanifold $D_i\cup D_f$.
In general, we have found it to be a difficult topological problem
to identify for which $D$-brane configuration $D_i\cup D_f$ and which choice 
of the
two-forms $\al_i$ and $\al_f$ the cocycle $(\o,\al_i,\al_f)$ is integer-valued.

Fortunately enough, if  the maximal compact subgroup of $D$ is simple
and simply connected,
we  have the key for solving our problem:
we draw the interacting $D$-brane diagrams directly in the double
and repeat the discussion of the section 2, using the duality invariant
first-order action (25). The action (25) also contains the WZNW term
but now  the forms $\alpha$ vanish. This means that the pairing
of the cocycle $(c,\alpha_F=0,\alpha_{Fd}=0)$
with any relative cycle $\gamma$ is just
\be \la (c,\alpha_F,\alpha_{Fd}),\gamma\ra=\int_{\gamma}c .\ee
Recall that we assumed $\pi_1(F)=0$. By the Hurewicz isomorphism,
we obtain $H^2(F)=0$ , hence every cycle in the relative
singular homology $H_3(R,D_i\cup D_f)$ can be represented by a cycle
in $H_3(R)$. This means that what matters is only whether $c$ is the standard
integer-valued three-cocycle in the third de Rham cohomology $H^3(D)$
of the Drinfeld double. But it is, because  $H^3(D)=H^3(K)$, where
$K$ is the simple simply connected maximal compact subgroup of $D$, and it 
is known that
the WZNW three-form restricted to $K$ is the integer-valued cocycle.

We find quite appealing that the path integral for the $D$-branes 
configurations seems to be topologically more easily tractable
by using the duality invariant formalism on the Drinfeld double.
On the other hand, an account of the local world-sheet phenomena,
like a short-distance behaviour, seems to be more difficult when
working with the non-manifestly Lorentzian
 first order Hamiltonian action (25). We plan to study this issue
in detail in a near future.

\section{Example: $SU(N)$ WZNW model}
 
Now we shall study examples of this general construction
of the self-dual WZNW models. Consider the group $SL(N,{\bf C})$ viewed as
the real group and the following invariant non-degenerate  bilinear form
on its algebra\footnote{The normalization of the bilinear form is always
such that the resulting action of the $SU(N)$ WZNW model will be properly
normalized  in order to meet the requirement that the WZNW three-form is
the integer-valued cocycle.}
\be \la X,Y\ra = {\rm Im}[(a^*)^2\T XY],  \qquad {\rm Im} a^2=4.\ee
The group $SL(N,\bc)$ equipped with the invariant bilinear form is the Drinfeld
double for every choice of the complex parameter $a$
satisfying the normalization constraint in (62). Two isotropic subalgebras
$\cg$ and $\cgt$ of $\cd$ are all upper and     lower triangular matrices
respectively with diagonal elements being $\lambda_k a, ~\lambda_k\in {\bf R}$
for $\cg$ and $\ti\lambda_k ia, ~\ti\lambda_k\in {\bf R}$ for $\cgt$.
Obviously, the index $k$ denotes the position on the diagonal and 
lambdas are constrained by the tracelessness condition. 

An example of $SL(2,\bc)$:
\be \cg=\left(\matrix{\lambda a&z\cr 0& -\lambda a}\right),
\quad \cgt=\left(\matrix{\ti\lambda ia&0\cr \ti z&-\ti\lambda ia}\right),\ee
where $z,\ti z$ are arbitrary complex numbers.

The dual pair of the $\si$-models is encoded in the choice of the 
half-dimensional subspace ${\cal R}$ of the Lie algebra $\cd$ of the double.
We choose ${\cal R}$ to be the $su(N)$ subalgebra of the algebra $sl(N,\bc)$.
Following our discussion above, it is easy to find the principal
fibrations of $SL(N,\bc)\equiv D$, corresponding to the algebras ${\cal R}$,
 $\cg$ and 
$\cgt$. The total space of the bundles is always the double $D$, the fibres
are $SU(N)$, $\exp{\cg}\equiv G$ and $\exp{\cgt}\equiv \ti G$ and the bases are
$SU(N)\backslash D$, $D/G$ and $D/\ti G$ respectively. Note that 
every fiber of  all three
fibrations can be obtained from the fiber crossing the unit element $e$ of $D$ 
by either the right (for $SU(N)$)
or the left (for $G$ and $\ti G$) action of some element of the double $D$.
In the particular example of the double $SL(N,\bc)$, the intersection
of a fibre $SU(N)$ with  fibres $G$ or $\ti G$ occurs always
precisely  at one point. 
It is not difficult to prove this fact. We already know that the intersection
always exists because $SU(N)$ is compact and $SL(N,\bc)$ is connected 
(cf. sec 3). 
If both fibers $SU(N)$ and $G$ (or $\ti G$) cross the unit element
of the double (which is the intersection point), it is obvious that a 
non-unit element of $G$ (or $\ti G$)
cannot be a unitary matrix. Thus the intersection is unique in this case.
Also an intersection $r$ of  the  $SU(N)$ fiber crossing
the unit element of $D$ with some fiber $G$ must 
be  unique. Indeed, the $G$ fiber can be then written as $rG$ where
$r\in SU(N)$. By the left action of $r^{-1}$ the $G$ fiber can be transported 
to the origin of $D$ where there is just one intersection.

Hence we conclude:  for our  data $D=SL(N,\bc),G,\ti G$ and ${\cal R}=su(N)$,
the both models of the dual pair (16) are the standard $SU(N)$ WZNW models,
because the restriction of the bilinear form (62) to ${\cal R}$ is
nothing but the standard Killing-Cartan form on $su(N)$.

Now   we may choose
the subgroup $F$ of $D$, which defines the $D$-branes in the
double, to be equal to $G$. Thus we have given a concrete meaning
to our  so far abstract construction.

It may be of some interest  to  provide few
explicit formulas for  the $SL(2,\bc)$ Drinfeld double.
The both cosets $D/G$ and $D/\ti G$ can be identified with the group
$SU(2)$. Recall that the space of all $D$-branes corresponding to the choice
$F$ is parametrized by the {\it left} coset $F\backslash D$. 
In our case $F\backslash D$ can also
be identified with $SU(2)$, hence a generic $D$-brane ($Fd$) in the
double is a set of $SL(2,\bc)$ matrices of the form
\be Fd \equiv \left(\matrix{e^{\lm a}&z\cr 0&e^{-\lm a}}\right)
 \left(\matrix{C&-E^*\cr E&C^*}\right),\ee
where $C,E$ are fixed complex numbers satisfying 
$CC^* +EE^*=1$, $\lm$ is a real
and $z$ a complex number.
In order to get
the $D$-branes in the cosets $D/G=D/F$ and $D/\ti G$, we have to project $Fd$
on $SU(2)$ along $G$ and $\ti G$, respectively:
\be Fd= \left(\matrix{A&-B^*\cr B&A^*}\right) 
\left(\matrix{e^{\eta a}&w\cr 0&e^
{-\eta a}}\right), \quad \eta\in {\bf R}, \quad w\in \bc ,\ee
\be Fd= \left(\matrix{\ti A&-\ti B^*\cr \ti B&\ti A^*}\right) 
\left(\matrix{e^{\ti\eta ia}
&0\cr \ti w&e^{-\ti\eta ia}}\right), \quad \ti\eta\in {\bf R},
\quad \ti w\in \bc .\ee
Here again 
$AA^*+BB^*=1$ and the same constraint is of course true also for $\ti A$ and 
$\ti B$. If $\lambda$ and $z$ vary then  $A$ and $B$ sweep a submanifold
of $SU(2)$, which is just the  $D$-brane in $R$ and $\ti A$ and $\ti B$ sweep
the dual $D$-brane in $R$.

There may occur three qualitatively different possibilities:

\noindent 1. Both $C$ and $E$ do not vanish (a generic case).

\noindent Then $B\neq 0$ and  it is convenient to parametrize $D$ and $B$ as
\be E=e^{E_1 a} e^{E_2 ia},\qquad B=e^{B_1 a} e^{B_2 ia},
\quad B_i,E_i\in{\bf R} .\ee
The original $D$-brane is then a two-dimensional submanifold of $SU(2)$
characterized by the condition
\be B_2=E_2.\ee
The dual $D$-brane is a three dimensional submanifold of $SU(2)$ which
is complement of the circle $A=0$.

\noindent 2. $C=0$.

\noindent The original $D$-brane is the same as in 1. but the
dual $D$-brane is just the one-dimensional circle $A=0$.

\noindent 3. $E=0$.

\noindent The original $D$-brane is a point $A=C$ and the dual $D$-brane
is the same as in 1.

It is not difficult to compute also the two-form $\alpha$ on the $D$-brane
(cf. (38)).
For doing this, we have just to know the mapping $k(r)$ from 
the original $D$-brane in $R$ to $G$
and the dual mapping $\ti k(r)$ from the dual $D$-brane in 
$R$ to $\ti G$ (cf. (35)). We do that
for the case 3 choice $C=1$ and $E=0$. The original map $k$ is trivial,
since the $D$ brane is just the point $A=1$ but the dual map $\ti k$ 
is nontrivial and it reads
\be \ti k(A,B)=\left(\matrix{e^{iA_2 a}&0\cr -e^{-A_1 a}B&e^{-iA_2 a}}\right).
\ee
Here
\be 0\neq A^*\equiv e^{A_1 a}e^{iA_2 a}, \quad A_1,A_2\in{\bf R}.\ee 
Now insert (69) in (38) and find
\be 8\pi\ti\alpha_f=
{\rm Im}[{a^*}^2\{(-BdB^*+B^*dB)a\w(A_1+idA_2)+
dB^*\w dB -2ia^2 dA_1\w dA_2\}].\ee
It is easy to compute the exterior derivative of $\ti\al_f$:
\be 8\pi d\ti\al_f=2{\rm Im}[{a^*}^2 a~ dB^*\w dB \w d(A_1+iA_2)]=
c_R\vert_{D_f}.\ee
In words: the exterior derivative of $\al_f$ is equal to the restriction
of the WZNW three-form $c_R$ on the $D$-brane $D_f$.  

It may be interesting to remark that in the case of the $SU(N)$ WZNW
models there are no topological obstructions in quantizing the
model on the topologically trivial open strip world-sheet. Thus, we do not
have to lift the $D$-brane configuration to the double in order to 
make the argument but we can directly proceed at the level of the
$D/G\equiv SU(N)$ target. 
Indeed,  choose two different
surfaces lying in the same $D$-brane 
and interpolating between the edges of the `diadem'. Their
oriented sum does not have a boundary and it is topologically the two-sphere.
If we happen to show that the second homotopy group $\pi_2(D_{i(f)})$ of
the $D$-brane $D_{i(f)}$ vanishes then the two interpolating surfaces are
homotopically equivalent and there is no ambiguity coming from the
WZNW term (cf. sec 2). 

It is easy to prove that $\pi_2(D_f)$
vanishes if the $D$-brane $D_f$ was obtained by our method of projecting
the isotropic surface $Fd$ from the double. Our basic tool
is the long exact homotopy sequence \cite{Schw}:
\be \pi_2(F)=0\to \pi_2(F/H)\to\pi_1(H)\to \pi_1(F)\to \pi_1(F/H)\to
\pi_0(H)\to 0=\pi_0(F),\ee
which holds for a connected group $F$ and its arbitrary subgroup
$H$;  note  that $\pi_2$ of any Lie
group vanishes. Now the $D$-brane on $SU(N)$ is gotten by projection
of the surface $Fd$ from the double to $D/G$. 
This means that topologically it can
be identified with the coset $F/dGd^{-1}\cap F\equiv F/H$.
We observe that in our $SL(N,\bc)$ context the group $F$ can be topologically
identified with its algebra $\cf$ because the usual exponential
mapping $\exp{\cf}=F$ is one-to-one. So it is one-to-one
for any its connected subgroup including the unity component of $H$. 
Hence $\pi_1(H)=0$ and since   $\pi_1(F)=0$,
from the sequence
(73) we conclude that $\pi_2$ of the $D$-brane in $SU(N)$ vanishes.
 
We should mention that from the exact sequence (73) it also follows 
that $\pi_1(D_f)=\pi_0(H)$. In general, for 
 our $SU(N)$ case the group $H$ is not connected.
This means that, strictly speaking, the diadem in our argument must
be equivalent to the zero element of $H_2(R,D_f)$, or, in other, words
it must be a relative two-boundary.
If the diadem is a non-trivial relative 
two-cycle we use the results of section 2
and  evaluate its contribution
by choosing the extension $\ti f:Z\to U$ of the homomorphism $f:B\to U$.

\end{document}